\documentclass[reprint,aps,prl,superscriptaddress]{revtex4-1}

\pdfoutput=1

\usepackage{graphicx}
\usepackage[dvipsnames]{xcolor}
\usepackage{calc} 

\usepackage{amsmath}
\usepackage{amssymb}
\usepackage{bm}

\usepackage{braket}
\usepackage{units}

\usepackage{bm}
\usepackage{eurosym}

\usepackage{url}
\usepackage[colorlinks=true,
        linkcolor=black,
        citecolor=black,
        filecolor=black,
        urlcolor=black,
        bookmarks=true,
        bookmarksopen=true,
        bookmarksopenlevel=3,
        plainpages=false,
        pdfpagelabels=true,hyperfootnotes,
        pdfauthor={C. Lang, D. Bozyigit, C. Eichler, L. Steffen, J. M. Fink, A. A. Abdumalikov Jr., M. Baur, S. Filipp, M. P. da Silva, A. Blais, and A. Wallraff},
        pdftitle={Observation of Resonant Photon Blockade at Microwave Frequencies using Correlation Function Measurements}
        ]{hyperref}

\newcommand{\graphicsDir}{./graphics}

\renewcommand{\nicefrac}[2]{#1/#2}

\begin{document}

\title{Observation of Resonant Photon Blockade at Microwave Frequencies\\using Correlation Function Measurements}

{\author{C.~Lang}
\author{D.~Bozyigit}
\author{C.~Eichler}
\author{L.~Steffen}
\author{J.~M.~Fink}
\author{A.~A.~Abdumalikov~Jr.}
\author{M.~Baur}
\author{S.~Filipp}
\affiliation{Department of Physics, ETH Z\"urich, CH-8093, Z\"urich, Switzerland.}
\author{M.~P.~da~Silva}
\author{A.~Blais}
\affiliation{D\'epartement de Physique, Universit\'e de Sherbrooke, Sherbrooke, Qu\'ebec, J1K 2R1 Canada.}
\author{A.~Wallraff}
\affiliation{Department of Physics, ETH Z\"urich, CH-8093, Z\"urich, Switzerland.}
\date{\today}

\begin{abstract}
Creating a train of single photons and monitoring its propagation and interaction is challenging in most physical systems, as photons generally interact very weakly with other systems. However, when confining microwave frequency photons in a transmission line resonator, effective photon-photon interactions can be mediated by qubits embedded in the resonator. Here, we observe the phenomenon of photon blockade through second-order correlation function measurements. The experiments clearly demonstrate antibunching in a continuously pumped source of single microwave photons measured using microwave beam splitters, linear amplifiers, and quadrature amplitude detectors.  We also investigate resonance fluorescence and Rayleigh scattering in Mollow-triplet-like spectra.
\end{abstract}

\maketitle


Sources of radiation differ not only by their frequency, but also by the statistical properties of the emitted photons~\cite{Walls1994}. Thermal sources emit radiation that is characterized by an enhanced probability of emitting photons in bunches. Coherent sources, such as a laser, emit radiation with a Poisson-distributed photon number. The statistics of these two sources can be explained classically. In contrast, individual atoms emit photons one by one well separated in time from each other, a phenomenon for which antibunching -- a unique quantum characteristic of the field -- can be observed.

In strongly nonlinear systems, a phenomenon known as photon blockade \cite{Tian1992a,Imamoglu1997} can be used to generate a train of single photons that displays antibunching. Photon blockade is usually realized in cavity quantum electrodynamics (QED) setups. Coherent radiation at the input of a cavity coupled to an anharmonic system, such as a single atom, is converted into a train of single photons in the transmitted light. The nonlinearity of the atom-cavity system prevents more than a single excitation of the same energy entering the cavity. Only once the photon has left the cavity can the system be reexcited, realizing a single-photon turnstile device. The transmitted radiation has two important characteristics: sub-Poissonian photon statistics and photon antibunching. On the one hand, sub-Poissonian statistics are experimentally demonstrated when the second order correlation function fulfills the inequality $g^{(2)}(\tau) \leq 1$ for all times $\tau$. On the other hand, photon antibunching is demonstrated by a rise of $g^{(2)}(\tau)$ with $\tau$ increasing from $0$ to larger values while $g^{(2)}(0) < g^{(2)}(\tau)$, as discussed in detail in Ref.~\onlinecite{Zou1990}.

At optical frequencies, resonant photon blockade -- cavity and atom share the same resonance frequency -- was demonstrated with a single trapped atom in an optical cavity~\cite{Birnbaum2005a}. These measurements suffer from adverse effects of trapping beams, micro-motion of the atom in its trap, and the necessity of post-selecting data for instances of single-atom measurements. In the solid-state, resonant photon blockade was demonstrated with a quantum dot in a photonic crystal cavity~\cite{Faraon2008}. Those experiments suffered from quantum dot blinking, and limited detector time resolution. Our experiments are done in the microwave regime with a single superconducting artificial atom resonantly coupled to a transmission line resonator, realizing a cavity QED setup~\cite{Haroche1992} in a circuit reaching the strong coupling limit~\cite{Wallraff2004b,Schoelkopf2008}. The artificial atom at rest, which is here well approximated by a two-level system, has a strong, fixed coupling to the resonator. In addition, our setup benefits from high-efficiency emission of photons in forward direction by employing an asymmetric quasi-one-dimensional resonator dominated by a single mode resonant with the artificial atom. This is in contrast to the atomic case for which the multi-mode structure of the cavity is important~\cite{Birnbaum2005a}. Also, the effective polarization of the radiation is fixed by the boundary conditions enforced by the superconducting metal forming the resonator, and thus does not play a role in our experiments.

In this paper we present correlation function measurements of continuous sources of single photons, coherent, and thermal radiation in the microwave frequency domain. In particular, we investigate the phenomenon of photon blockade both in resonance fluorescence and second-order correlation function measurements, displaying sub-Poissonian photon statistics and antibunching. Photon blockade in superconducting circuits has also been independently studied in the dispersive regime in Ref.~\onlinecite{Hoffman2010}.

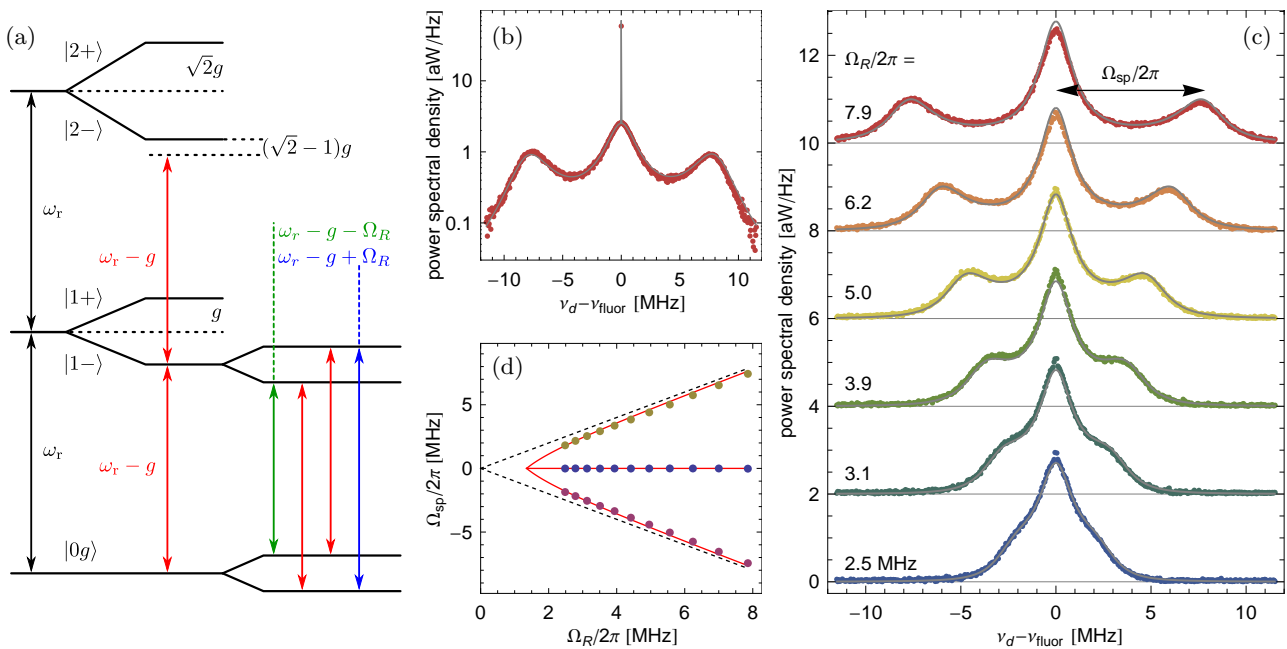
\begin{figure*}
  \centering
  \includegraphics[scale=1.0]{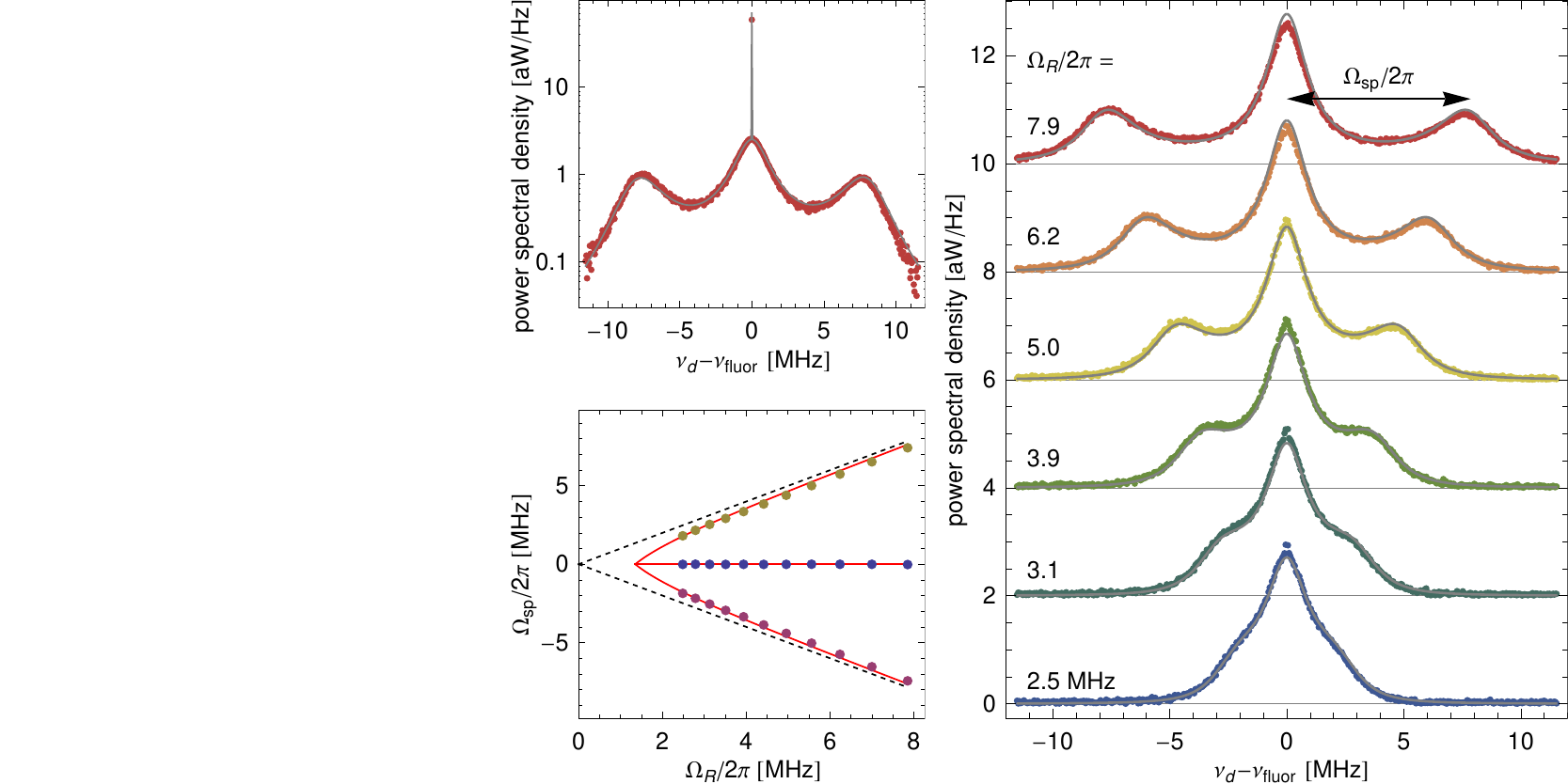}
  \begin{picture}(0,0)
	\put(-485, 20){\includegraphics[scale=1.4]{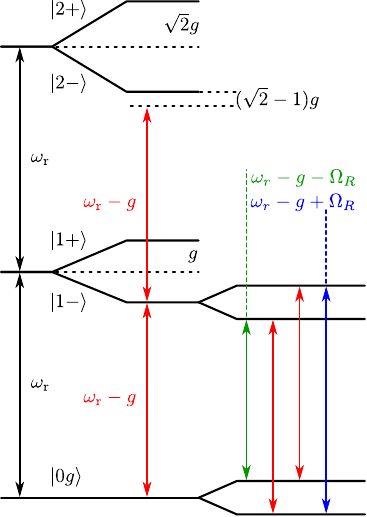}}
  \end{picture}                    
  \begin{picture}(0,0)             
	\put(-490, 228){(a)}  
	\put(-306, 228){(b)}  
	\put( -22, 228){(c)}  
	\put(-306, 103){(d)}  
  \end{picture}
  \caption{{\textbf{Rayleigh scattering and resonance fluorescence of lower Jaynes-Cummings doublet.}}
	(a)~Energy level diagram of a resonantly coupled cavity QED system driven with amplitude $\Omega_R$ on the ground state $|g0\rangle$ to lower doublet $|1-\rangle$ transition. The Mollow-type transitions arising from the dressing of the dressed states by the strong drive are also indicated on the side.
	(b)~Measured resonance fluorescence spectrum including Rayleigh scattering peak (dots) at fixed drive amplitude of $\nicefrac{\Omega_R}{2\pi} = \unit[7.9]{MHz}$ and simulated spectrum (solid line).
	(c)~Measured resonance fluorescence spectrum vs. (indicated) drive amplitude $\nicefrac{\Omega_R}{2\pi}$ (dots) and analytical spectrum (solid lines). The Rayleigh peak has been omitted in these plots.
	(d)~Measured Mollow side peak frequencies $\Omega_{\rm{sp}}$ \textsl{vs.}~drive amplitude $\Omega_R$ (dots), linear dependence $\Omega_{\rm{sp}}=\Omega_R$ (dashed black lines) and calculated frequencies $\Omega_{\rm{sp}}$ (solid red lines) are shown.}
  \label{fig:MollowTriplett}
\end{figure*}
Our experimental setup is composed of two essential ingredients: a photon source and a quadrature amplitude detection system from which we extract the photon statistics. The continuous single-photon source consists of a single superconducting artificial atom -- a transmon qubit~\cite{Koch2007} with transition frequency $\omega_a$ -- resonantly coupled to a transmission line resonator with resonance frequency
$\omega_r/2\pi = \omega_a/2\pi = \unit[6.769]{GHz}$. 
In this device the coherent dipole coupling strength
$\nicefrac{g}{2\pi} = \unit[73]{MHz}$ 
dominates over the dissipation due to photon loss from the cavity at rate
$\nicefrac{\kappa}{2\pi} \approx \unit[4]{MHz}$ 
and the qubit decay at rate
$\nicefrac{\gamma}{2\pi} \approx \unit[0.4]{MHz}$. 
When radiation impinges on the resonator input at frequency
$\nicefrac{(\omega_r-g)}{2\pi}$, 
only a single photon can enter at a time, see Fig.~\ref{fig:MollowTriplett}a. Additional photons are prevented from entering the resonator, as transitions into higher excited states are blocked due to the strong non-linearity of the resonantly coupled qubit-resonator system~\cite{Schuster2008,Hofheinz2008,Fink2008}. In analogy with measurements in mesoscopic systems, where electron transport is blocked by the strong Coulomb interaction in a confined structure, this process is called \emph{photon blockade}~\cite{Imamoglu1997}. Only once the photon has left the cavity can the next photon enter into the resonator, realizing a source of single photons.

In order to investigate the statistical properties of our microwave frequency radiation source, we have realized a scheme for measuring photon correlation functions using linear detectors~\cite{daSilva2010,Bozyigit2010a,Menzel2010} instead of single-photon counters which are still under development in this frequency domain~\cite{Chen2010b,Romero2009}. In our scheme, the radiation of the source is passed through an on-chip 50/50 beam splitter, then the signal in each output of the beam splitter is amplified using independent phase preserving linear amplifiers with system noise temperature
$T_n = \unit[10.6]{K}$. 
Finally, both quadrature amplitudes of each output signal are extracted in a heterodyne measurement similar to the one discussed in Ref.~\onlinecite{Bozyigit2010a}. Expectation values of field amplitude, power, first- and second-order correlation functions can be extracted from the instantaneous values of the measured quadrature amplitudes. We refer to Ref.~\onlinecite{daSilva2010} for a detailed theoretical discussion.

We set up our continuous single-photon source by tuning the transmon qubit transition frequency $\omega_a$ into resonance with the resonator using magnetic flux~\cite{Koch2007,Schreier2008}.  When probing the resonator transmission with a weak coherent tone resulting in an average resonator photon number $\braket{n} \ll 1$, we observe a characteristic vacuum Rabi mode splitting~\cite{Wallraff2004b,Fink2008} resulting from the anharmonic level structure shown schematically in Fig.~\ref{fig:MollowTriplett}a. In many experiments of this type, only the Rayleigh scattered (elastic and coherent) part of the transmitted amplitude is detected in a heterodyne measurement with a small effective bandwidth of $\sim \unit[50]{kHz}$. 
Here however, we have digitally recorded the resulting fields \textsl{vs.}~time in both arms of the beam splitter with a bandwidth of $\sim\unit[50]{MHz}$.
Instantaneous power spectra of the source are then calculated as the product of the Fourier transform of the time-dependent signals in each arm, which are subsequently averaged. Here, we observe not only the Rayleigh scattered radiation (narrow high-amplitude peak in Fig.~\ref{fig:MollowTriplett}b) but also the incoherently scattered resonance fluorescence part of the spectrum (broad low-amplitude triplet in Fig.~\ref{fig:MollowTriplett}b). The resonance fluorescence spectrum is characterized by three spectral lines (four transitions [Fig.~\ref{fig:MollowTriplett}a], two of which are degenerate) forming a Mollow triplet of a resonantly driven effective two-level system. The two levels are  realized by the joint ground state $\ket{g0}$ and the lower energy state of the first doublet $\ket{1-} = (\ket{g1} -\ket{e0})/\sqrt{2}$ of the Jaynes-Cummings ladder. The dressing of these dressed states by the drive field has been discussed theoretically in Ref.~\onlinecite{Tian1992a} and has also been experimentally investigated with superconducting circuits considering only the Rayleigh scattered part of the radiation~\cite{Bishop2009a}.

The full spectrum is in excellent agreement with the numerically calculated steady-state solution of the master equation taking into account two qubit levels and five resonator levels (solid line in Fig.~\ref{fig:MollowTriplett}b). For this calculation, we use the device parameters quoted above and take into account the finite bandwidth of our detection system. Also, the analytically calculated fluorescence spectrum of the coherently driven effective two-level system (solid lines in Fig.~\ref{fig:MollowTriplett}c) is virtually indistinguishable from the master equation calculation and the data. Here we include dephasing and do not make approximations for the strength of the drive~\cite{Carmichael1999Book}. To correctly capture the amplitude of the coherently scattered radiation in the analytical calculation the higher doublet $\ket{2-}$ needs to be included.

The frequency $\Omega_{\rm{sp}}$ by which the Mollow side peaks are offset from the central peak is observed to depend on the drive amplitude $\Omega_R$ (Fig.~\ref{fig:MollowTriplett}c). Lorentzian fits to the triplet spectrum yield $\Omega_{\rm{sp}}$, showing an approximately linear scaling with $\Omega_R$ at large drive amplitudes, see Fig.~\ref{fig:MollowTriplett}d. However, $\Omega_{\rm{sp}}$ is significantly smaller than $\Omega_R$ at drive strengths smaller than the characteristic rate of dissipation, an effect that is accurately explained by the analytical two-level model (see red solid lines in Fig.~\ref{fig:MollowTriplett}d)~\cite{Carmichael1999Book}. Similar Mollow triplet-like structures have also been observed in strongly driven superconducting flux and charge qubits using different detection techniques~\cite{Baur2009,Sillanpaa2009,Astafiev2010}.

We note that, for these measurements, the uncorrelated noise added by the two independent amplifiers is efficiently averaged out~\cite{Agarwal1994}, and the residual noise offset -- a factor of $10^3$ 
smaller than the noise introduced by a single amplifier -- is determined by performing a reference measurement where the system is left in the ground state and then subtracted from the data~\cite{daSilva2010}. 

The experiments discussed above clearly demonstrate the resonance fluorescence emitted from the cavity when it is weakly driven on the lower Rabi resonance $(\omega_r-g)$. In this limit, photon blockade is expected to be observable in measurements of the normalized second-order correlation function $g^{(2)}(\tau)$. We extract $g^{(2)}(\tau)$ from a measurement of the cross-correlation of the power detected between the two outputs of the 50/50 beam splitter~\cite{daSilva2010}. The constant offset due to the noise added by the amplifiers is subtracted and the correlation function is normalized to unity for times $\tau \rightarrow \infty$. At low drive amplitudes ($\nicefrac{\Omega_R}{2\pi} = \unit[2.5]{MHz}$), we observe sub-Poissonian photon statistics characterized by $g^{(2)}(\tau) \leq g^{(2)}(\infty)$, which is fulfilled within the experimental noise. Additionally, photon antibunching is observed since $g^{(2)}(\tau)$ rises for $\tau$ away from $\tau=0$ while $g^{(2)}(0)$ is at a minimum (Fig.~\ref{fig:Correlations}a). For $\tau \rightarrow \infty$ we note that $g^{(2)}$ approaches a constant value, as expected. We observe a small overshoot of $g^{(2)}(\tau)$ at around $\tau = \nicefrac{\pi}{\Omega_R} = \unit[200]{ns}$ in Fig.~\ref{fig:Correlations}a. This indicates a correlation between a photon emitted at time $t$ and a second photon emitted with high probability at the later time $\nicefrac{(t+\tau)}{\Omega_R} = \pi$ at which the drive has coherently re-excited the coupled system.

\begin{figure}[t!]
\centering
\includegraphics[trim=0 0 0 0,clip]{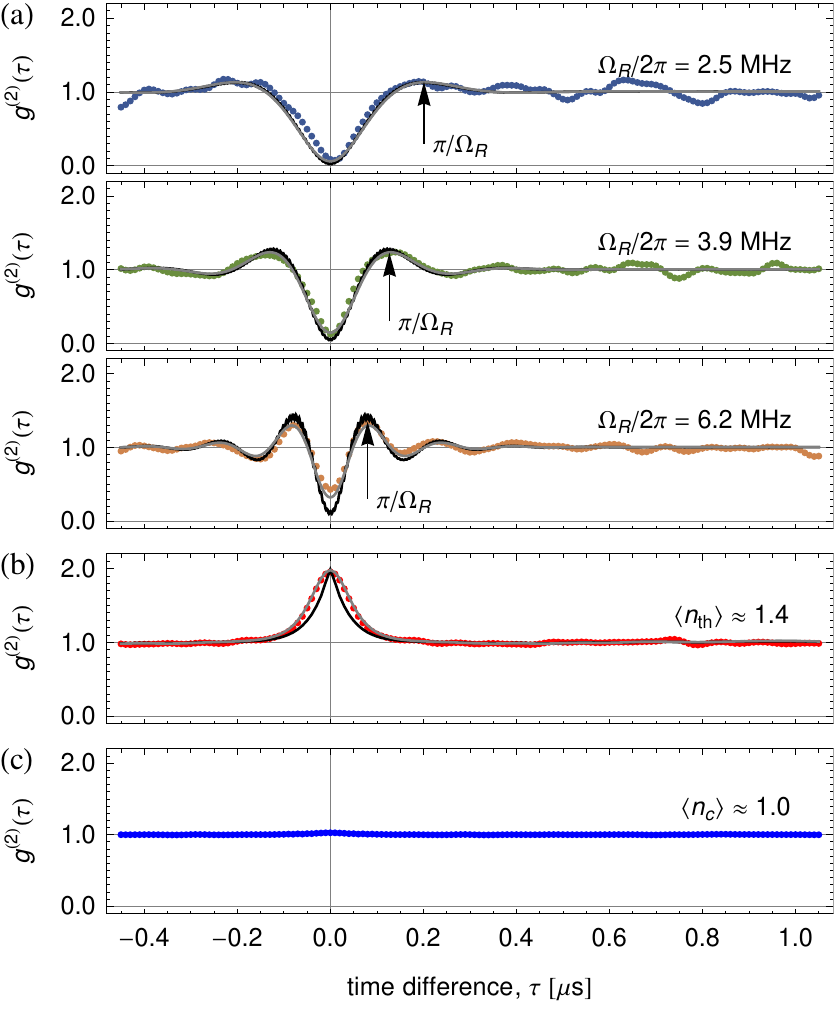}%
\caption{\textbf{Correlation function measurements.}
  (a)~Second-order correlation function measurements $g^{(2)}(\tau)$ (dots) for indicated drive amplitudes $\Omega_R$ and master equation calculation with and without accounting for finite measurement bandwidth (gray and black lines, respectively).
  (b)~$g^{(2)}(\tau)$ for a thermal field with mean photon number $\langle n_{\rm{th}}\rangle\sim1.4$ in the resonator.
  (c)~$g^{(2)}(\tau)$ for a coherent drive with $\langle n_{\rm{c}}\rangle\sim1$.
  }\label{fig:Correlations}
\end{figure}

When increasing the drive amplitude $\Omega_R$ we observe characteristic oscillations in the measured $g^{(2)}(\tau)$ exactly at the frequency $\Omega_R$ (Fig.~\ref{fig:Correlations}a) while we continue to clearly observe the signature of antibunching at $\tau = 0$. We quantitatively compare the measured data to numerical calculations of $g^{(2)}(\tau)$ (see black lines in Fig.~\ref{fig:Correlations}a) based on a master equation calculation using the known system parameters. Considering the finite bandwidth $\sim \unit[20]{MHz}$ of the digital filter used in the quadrature data acquisition, we find excellent agreement between the measured data and the calculations, see gray lines in Fig.~\ref{fig:Correlations}a.
The small residual deviations of the measured $g^{(2)}(\tau)$ from the simulations are due to the noise added by the amplifiers.
We note that each data trace was collected over $17$ hours 
corresponding to approximately $5.5\times10^{10}$ 
measured photons and $\unit[15.75]{Tbyte}$ 
of analyzed quadrature amplitude data using fast field-programmable gate array based electronics~\cite{Bozyigit2010a}. The presented data clearly demonstrates the phenomenon of photon blockade in the microwave domain detected using second-order correlation function measurements.

For reference we have also measured $g^{(2)}(\tau)$ when populating the resonator with a mean thermal photon number $\braket{n_\text{th}} \approx 1.4$.
The quasi-thermal field distribution was realized by mixing a fixed frequency microwave tone with a large bandwidth white noise source~\cite{Fink2010}. We clearly observe bunching $g^{(2)}(0) = 2$ of the thermal radiation emitted from the resonator. $g^{(2)}(\tau)$ approaches unity on the time scale of the cavity decay rate $\nicefrac{\kappa}{2\pi}$ also considering the finite detection bandwidth (gray line in Fig.~\ref{fig:Correlations}b). Performing a similar experiment with a coherent source derived from a strongly attenuated commercial microwave generator populating the resonator with $\braket{n_\text{c}} \approx 1.0$, we find $g^{(2)}(\tau) = 1$ everywhere, which is in good agreement with the temporal statistics of a coherent source.


We have performed correlation function measurements with linear quadrature amplitude detectors in the microwave frequency domain demonstrating photon blockade in a circuit QED system. We have also shown bunching of thermal photons and probed the second-order correlation function of coherent radiation.
The techniques and results presented in this paper have the potential to inspire new work controlling the flow of photons, generating and detecting individual photons and investigating single-photon effects in superconducting circuits. In particular, the observation of photon blockade will enable future experimental work on photon interactions in cavity arrays that is actively theoretically investigated~\cite{Schmidt2010a,Hartmann2008,Koch2009a,Angelakis2007,Greentree2006}.

\begin{acknowledgments}
The authors would like to acknowledge fruitful discussions with Barry Sanders. This work was supported by the European Research Council (ERC) through a Starting Grant and by ETHZ. M.P.d.S.~was supported by a NSERC postdoctoral fellowship. A.B.~was supported by NSERC, CIFAR, and the Alfred P.~Sloan Foundation.
\end{acknowledgments}



\bibliographystyle{apsrev4-1_4auth}
\bibliography{../QudevRefDB.bib}

%

\end{document}